\documentclass[aps,prl,showpacs,amssymb, amsmath,nobibnotes,a4paper, 10pt,twocolumn]{revtex4-2}

\usepackage[dvips]{epsfig}
\usepackage{xcolor,soul}
\usepackage{graphics,graphicx}
\usepackage{epstopdf}
\usepackage{hyperref}
\usepackage{mhchem}
\usepackage{bm}

\DeclareGraphicsExtensions{.jpg,.pdf,.mps,.png}

\begin{document}

\title{Heating rate in a linear quadrupole trap} 
\author{A. Poindron, J. Pedregosa-Gutierrez, C. Champenois}
\affiliation{Aix-Marseille Université, CNRS, PIIM, Marseille, France}
\date{\today}
\vspace{-12em}
\begin{abstract}
In radio-frequency trap, the temperature of ion ensembles converges towards a hot equilibrium due to radio-frequency heating. This effect is detrimental to the stability of trapped ensembles and is the justification of cooling. The intensity of this radio-frequency heating increases with the amplitude of the radio-frequency field $q_x$. Using an analytical empirical formula, we show that the lifetime of the ion ensemble $t_0$ under cold condition increases with $q_x$ according to a power law $t_0\propto q_x^A$, and does not vary significantly for the several ion quantities $N$ tested. The temperature of the explosive onset $B$ decreases linearly with $q_x$. We also show that non-linear instabilities due to trapping parameters decreases $t_0$ and $B$, and produce a local increase of heating rate for certain temperature ranges.
\end{abstract}

\maketitle 

\section{Introduction}

For the investigation of constituents of matter and their dynamical behaviour, particle traps are an essential tool \cite{mihalcea2023}. They are appreciated for their ability to isolate a small amount of particles in a limited volume of space where observations and interactions can be carried out. For the trapping of charged particles, the use of radio-frequency (rf) fields is very convenient. Nevertheless it introduces a forced motion in the ions dynamics which leads to radio-frequency heating \cite{nam17}, an effect detrimental to the stability of trapped ensembles and heating switching them from a cold equilibrium to a hot one \cite{poindron23}. Laser or sympathetic cooling can be used in order to bring the trapped ions to a low temperature equilibrium, limiting the effect of rf heating to a minimum, making ensembles of cooled ions suitable for many experiments where a high level of precision is required such as in : micro-wave \cite{Prestage91} and tera-hertz \cite{Champenois07} metrology, high precision spectroscopy \cite{morigi15}, quantum information experiments \cite{Cirac95,Blatt08,mokhberi20,willitsch20}, manipulation of cold molecular ions \cite{Roth05,willitsch08r,ColdMol09} or plasma stopping power experiments \cite{Bussmann06}. Conversely, we recently identified a new application for an ensemble of laser-cooled trapped ions as a detector for giant molecular ions \cite{poindron21} where rf heating can be exploited as a beneficial effect. In this principle, a molecular ion to be detected is introduced in a laser-cooled trapped Ca+ ensemble, and triggers a modification of the temperature of the ion ensemble, observable in its fluorescence. The radio-frequency heating is used as an amplifier for the initial perturbation that would be unable to induce a significant change in the ion ensemble temperature alone. In both cases an analytical description of rf heating would be beneficial for the design and operation of trapped charged particles experiments but remains unavailable. In this article we propose to study the temporal dynamics of rf heating through numerical simulations and aim at providing an analytical expression for the thermal evolution as well as the ensemble cold lifetime as a function of the rf amplitude.
Previous numerical works have been devoted to the study of the thermal evolution of trapped ensembles of charged particles. First studies pointing out the relation between rf heating and order-chaos transition were carried out around year 1990 \cite{hoffnagle88,Blumel89,Brewer90} for only two-ion systems. In 1991 the thermal evolution of ensembles up to 512 ions has been studied \cite{prestage91rf} during few hundreds of secular period, limiting the study to temperature below 50 mK and to rf amplitude at the limit of stability. In 2005, periodic edge conditions have been used to study the thermal evolution at longer time scales \cite{Ryjkov05}, reaching 10K, but in that case the study does not account for finite size effects which are non negligible \cite{Schiffer02,Schiffer02conf}. In \cite{Nam14,nam17} the rf heating are computed using an equilibrium forced by a friction, thus without considering the dynamics. For this article we propose to study the time-dependence of the temperature for ion ensembles up to 1024 ions and temperatures up to 400 K. We propose an analytical, empirical, function that can predict the cold lifetime of the ion ensemble in the cold equilibrium for three ensemble populations and several rf amplitudes. The knowledge of this cold lifetime is useful for the design and operation of experiments where the cooling may be deactivated occasionally. Also, this analytical function may allow to determine a threshold temperature whose knowledge may help design experiments such as proposed in \cite{poindron21}.

\section{Linear quadrupole trap}
\subsection{Trapping potential and motion equation}
For those simulations, the trap parameters are the same as in our previous article \cite{poindron23}. In a quadrupole linear Paul trap supplied with a radio-frequency voltage $U_{rf}$ with angular frequency $\Omega = 2\times 2\pi$ MHz and a static voltage $U_{st}$ the potential within its internal radius $r_0$ is expressed as
\begin{equation}
\Phi(x,y,t) = \frac{U_{st} + U_{rf}\cos\Omega t}{r_0^2}\left(x^2 - y^2 \right)
\end{equation}
with inner radius $r_0=2.5$ mm and $x$ and $y$ the two radial axis \cite{poindron21}. 

The dynamics of an ion $i$ with mass $m$ in this potential is described by the following equations :
\begin{align}
&\ddot{u}_i + \left( -\frac{\omega_z^2}{2} \pm \frac{2 Q U_{RF} \cos{\Omega t}}{m r_0^2} \right)u_i \notag \\
&\quad\quad\quad\quad = \frac{Q^2 }{4\pi \epsilon_0 m}\sum_{j=1,j\neq i}^N{\frac{u_i - u_{j}}{|\bm{r}_{i}-\bm{r}_{j}|^3}}  \label{eq:mathieu_r}\\
&\ddot{z}_i + \omega_z^2 z_i = \frac{Q^2 }{4\pi \epsilon_0 m}\sum_{j=1,j\neq i}^N{\frac{z_i - z_{j}}{|\bm{r}_{i}-\bm{r}_{j}|^3}} \label{eq:mathieu_z}
\end{align}
where $u_i$ stands for $x$ or $y$, and $\bm{r}_i = (x_i,y_i,z_i)$, $Q = z_e\times q_e = z_e 1.6\cdot 10^{-19}$ C the particle charge, $k_C=1/(4\pi\varepsilon_0)$ and $\varepsilon_0$ the permittivity. $a_u$ and $q_u$ are referred as stability parameters and defined as
\begin{align}
a_x &= -a_y = \frac{8QU_{st}}{mr_0^2\Omega^2} \label{eq:a_u} \\
q_x &= -q_y = \frac{4QU_{rf}}{mr_0^2\Omega^2}. \label{eq:q_u}
\end{align}
In the case of a single ion trapped, the right-hand term of equation \ref{eq:mathieu_r} cancels and can be recast into the so-called Mathieu equation. In this case the solution can be analytically derived and its stability determined as a function of the parameters $a_u$ and $q_u$.

The confinement along the trap axis, z-axis, is done with a static potential $U_G = m\omega_z^2z^2$ applied to the ends of each rod, generating a quadratic potential in our segmented trap \cite{poindron21}, characterised by the frequency $\omega_z/2\pi = 100$ kHz for a static potential applied of $U_{DC}=1$ V.

The secular frequencies are computed in the same manner as in \cite{poindron23}, and the trapping potential are chosen in order to obtain ion ensembles with unity aspect ratios.


\subsection{Temperature and heating rate}
The temperature of the ion ensemble is related to the velocity $v_i$ of the ions by the formula \cite{marciante10}
\begin{equation}
\frac{3}{2}k_BT = \frac{1}{2}\frac{m}{N} \sum_{i=1}^N \overline{v_i}^2 =    \langle \overline{v_i}^2 \rangle = \frac{1}{2}\frac{m}{N}\langle V_i^2 \rangle
\end{equation}
with $N$ the number of ions in the cloud, $k_B$ the Boltzmann's constant, and $\overline{v_i}$ the velocity of the ion $i$ over one rf period $\tau_{rf} = 500$ ns. Averaging over one radio-frequency period allows to smooth out the micromotion induced by the radio-frequency field, and proves to provide a satisfying averaged velocity for the measurement of the ensemble temperature \citep{Schiffer00}.

\subsection{Radio-frequency heating}

Because of radio-frequency heating, the temperature of an ensemble of charged particles increase over time towards a high temperature equilibrium. Radio-frequency heating is due the coupling introduced by the Coulomb forces, allowing energy from the radio-frequency field to be transferred to the Brownian motion of the ions \cite{Blumel89,prestage91rf,Ryjkov05}. One feature of this radio-frequency heating is the so-called fast transition from low temperature to high temperature equilibrium \cite{poindron23}, refereed as an "explosive onset" \cite{hoffnagle88}. Heating rate $H$ is defined as the temperature variation for a given unit of time. Because temperature is computed so the micromotion is ignored, heating rate can be understood as the time derivative of the temperature $H=\mathrm{d}T/\mathrm{d}t$.

\section{Molecular dynamics simulations}

\begin{figure*}[bth]
  \includegraphics[width=0.70\textwidth]{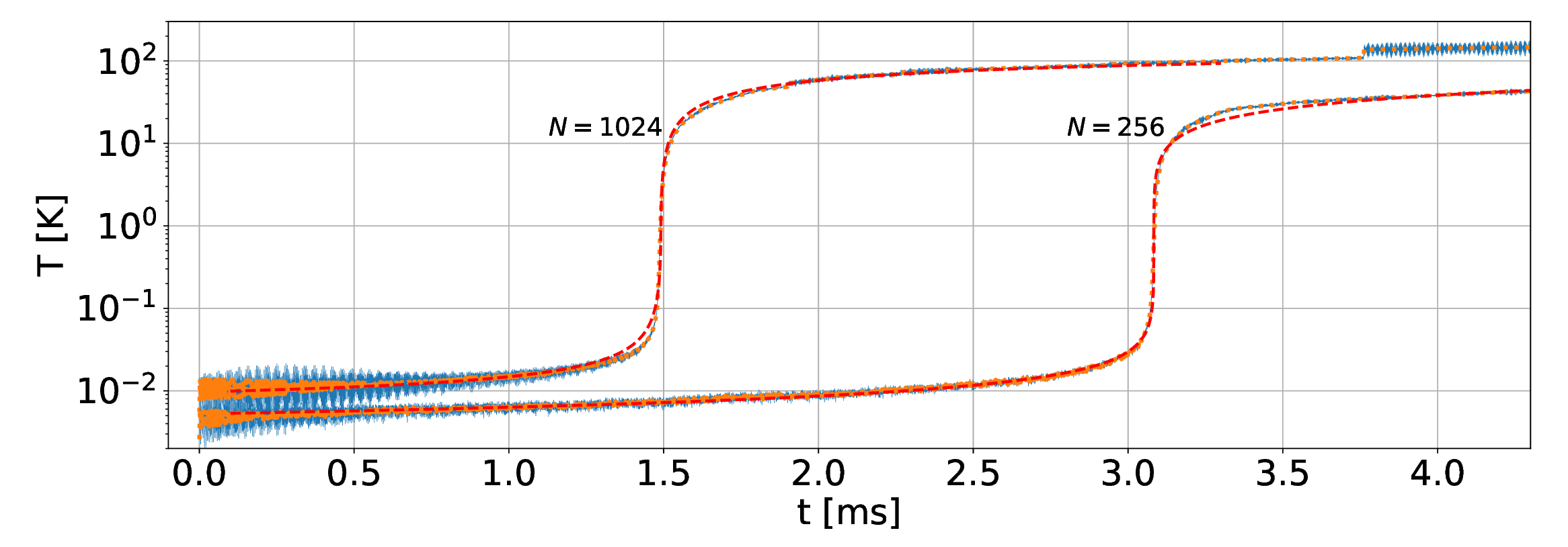}
  \caption{Temperature evolution for $q_x=0.64$, $U_{DC} = 15.4$ V. The blue line is the temperature computed from the cycle average velocity of ions, the dotted orange line is the temperature filtered with a 8th-order digital Butterworth filter forward-reverse in time \cite{lyle14}, the dashed red line is the analytical fit of the temperature with function $\psi$.}
  \label{fig:rf_curve}
\end{figure*}

\subsection{Parameters}

The numerical simulations are carried out with a molecular dynamics program based on velocity-Verlet algorithm \cite{poindron21}, integrating Coulomb repulsion and electric trapping forces, but neither laser cooling nor friction. Ensemble of $N=[256,512,1024]$ ions are simulated for each of the set of trapping parameters in table \ref{tab:params}, initially for a total of 27 different conditions. Some additional condition have been tested to refine the results when required. The trapping voltages are chosen so the trapping potential is spherical.

\begin{table*}
\centering
\begin{tabular}{|c|c|c|c|c|c|}
\hline 
$q_x$ & $0.54$ & $0.56$ & $0.58$ & $0.60$ & $0.62$ \\ 
\hline 
$a_z$ & $5.55\cdot10^{-2}$ & $6.04\cdot10^{-2}$ & $6.56\cdot10^{-2}$ & $7.12\cdot10^{-2}$ & $7.71\cdot10^{-2}$ \\
\hline
\hline 
$q_x$ & $0.64$ & $0.66$ & $0.68$ & $0.70$ & \\ 
\hline 
$a_z$ & $8.35\cdot10^{-2}$ & $9.04\cdot10^{-2}$ & $9.77\cdot10^{-2}$ & $1.05\cdot10^{-1}$ & \\ 
\hline 
\end{tabular} 
\caption{Trapping parameters for the first systematical set of simulations.}
\label{tab:params}
\end{table*}

While the simulation time is expected to increase with $N$, it turns out $q_x$ is the main parameter affecting wall time, because the smaller $q_x$ the longer the simulation. In order to keep the simulations under two hours, no simulation is done with $q_x<0.54$ and $N>1024$. Additional conditions were rendered to study the behaviour of the system around $q_x=0.62$ where an instability is found \cite{drakoudis06}. The simulation time step $\tau_{simu}$ is taken as the highest value for which no significant change is visible when it is taken smaller. In this article the time step is taken so there are $n_t = 1000$ time step in one radio-frequency cycle $\tau_{simu} = 2\pi/(n_t\Omega)$.

\subsection{Analytical empirical formula}
The sigmoid shape of the temperature evolution in semilog-y representation (Fig. \ref{fig:rf_curve}) inspired the choice for the analytical function used to fit the data. Elementary logistic or arctangent functions have been tested, but showed a poor ability to fit the temperature evolution, which translated into inconsistent determinations of cold lifetimes as a function of the rf amplitude. A more general sigmoid function with 5 parameters is used in this article (Eq. \ref{eq:sigmoid}) \cite{Dunning15}, as follows :
\begin{equation}
\psi = A\frac{t-t_0}{\left( C + |t-t_0|^\gamma \right)^{1/\gamma}} + B \label{eq:sigmoid}
\end{equation}
with $A$ the amplitude of the function and $B$ the value of $\psi$ at $t=t_0$. $C$ and $\gamma$ are parameters related so $A/(C^{-1/\gamma})$ is the value of $\mathrm{d}\psi / \mathrm{d}t$ at $t=t_0$. $\mathrm{d}\psi / \mathrm{d}t$ is maximum for $t=t_0$. This function exhibits a fast transition between two levels with an inflection point of coordinates $(t_0,B)$. $t_0$ is the timing of the inflection point, related to the timing of the explosive onset in this article, and $B=T(t=t_0)$ the temperature at which this explosive onset occurs.

Because the temperature exhibits fast oscillations, unrelated to radio-frequency and secular motion, it is filtered before applying the fit. The filter is an 8th-order digital Butterworth low-pass filter, forward-reverse in time to ensure no time shift is induced \cite{lyle14} as with a Savitzky-Golay for instance. The cut frequency of this filter is chosen equal to the double of the highest of the first order secular frequencies ($\omega_x, \omega_y, \omega_z$).

It is not satisfying to use $t_0$ directly as the timing of the explosive onset because the initial temperature varies with $N$ for a fixed set of numerical parameters. A reliable cold lifetime $\Delta t$ should only depend on the initial temperature when its dependency regarding the initial temperature would be studied. In the case of this article, the cold lifetimes are studied as a function of the trapping parameters and the population. The cold lifetimes are then computed by measuring the time difference between the two following events : the ensemble reaches $T_{threshold} = 10$ mK and the explosive onset. This can be written as $\Delta t = t_0 - t(T = T_{threshold})$.

\subsection{Heating rate}

Radio-frequency heating is a phenomenon significantly occurring in long-time scale with respect to the radio-frequency period. Rather than carrying on a time derivate of the temperature sampled at the radio-frequency period it is more relevant to derivate the heating rate from a window-averaged temperature. Then the temperature is first averaged over a given number of radio-frequency periods $n_t$ before time derivation is applied. Furthermore, because the radio-frequency sampled temperature exhibits significant oscillation at short time scale unrelated to radio-frequency heating, averaging ensures that those oscillations are neglected with respect to the long term variations due to radio-frequency heating.

\section{Results}

\subsection{Temperature evolution}

Temperature evolution of trapped ensembles exhibits a bistable behaviour as described in \cite{poindron23}. Figure \ref{fig:rf_curve} shows two temperature evolutions for two ensembles sizes and similar trapping parameters. Ions are initialized below $10^{-2}$ K before the explosive onset occurs within few milliseconds. The temperature of the ensembles reaches $100$ K after a delay similar to the delay between $10^{-2}$ and $1$ K.

\subsection{Cold lifetime}
\begin{figure}[h]
  \includegraphics[width=0.44\textwidth]{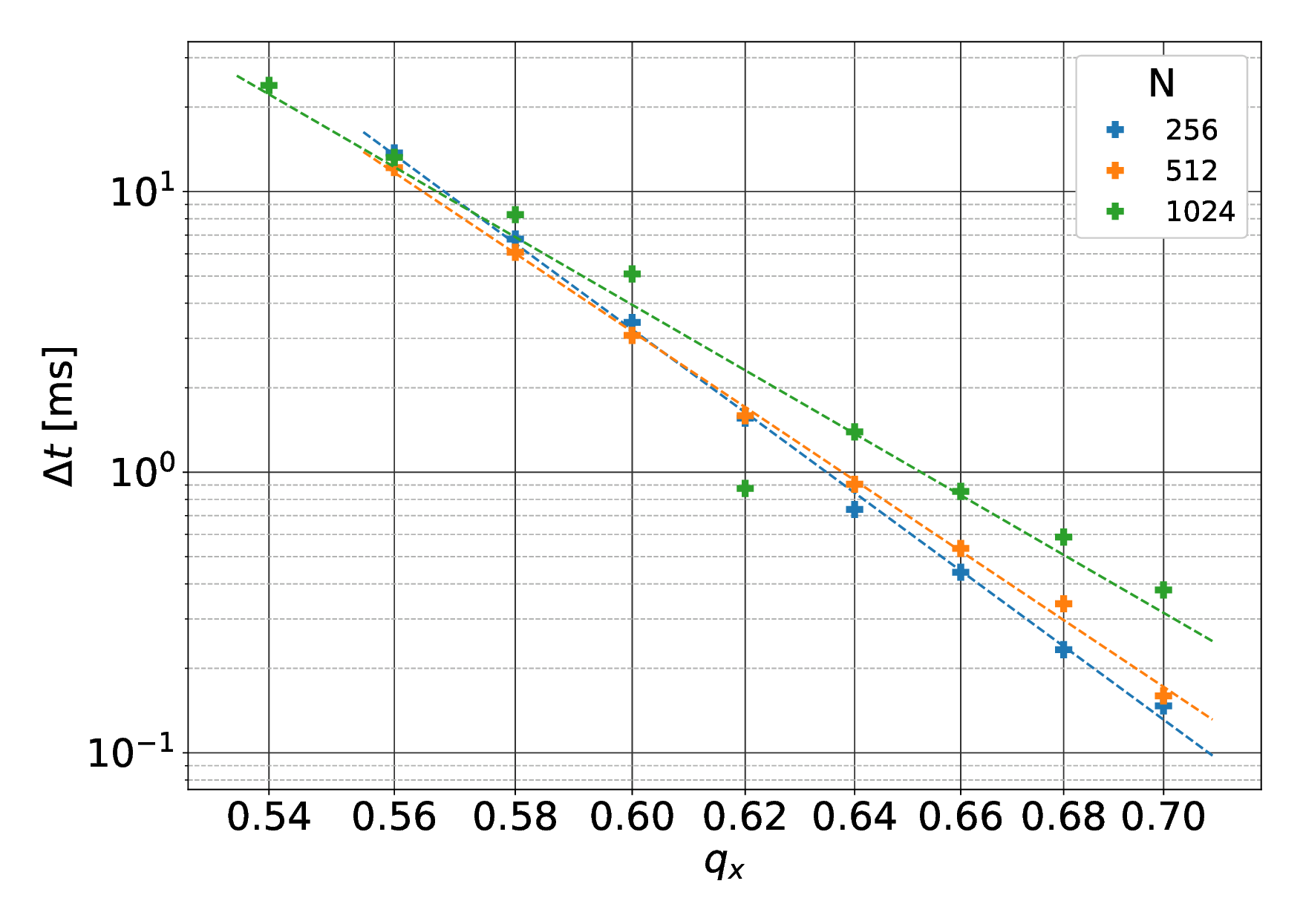}
  \caption{Ensemble lifetime $t_0$ as a function of Mathieu parameter $q_x$. The dashed lines are affine fits to the corresponding data (colour code).}
  \label{fig:t0vsqx}
\end{figure}

All temperature evolutions from simulated conditions were fitted with the function $\psi$. The ensembles lifetime $\Delta t$ for each condition are displayed in Fig. \ref{fig:t0vsqx} against the Mathieu parameter $q_x$. For each ensemble, $\Delta t$ vs. $q_x$ exhibits an affine behaviour in a loglog representation. This data is fitted with an affine function $\log \Delta t = p\log q_x + q$ which allows to express $\Delta t$ as a power law of $q_x$ so that $\Delta t = r q_x^p$. Values of $p$ and $r=10^q$ are presented in table \ref{tab:t0vsqx_fit}. It is not guaranteed the linear fit can extrapolate beyond the simulated range of parameters, but it already provides an idea of ensembles lifetimes. In particular, looking at the range from $q_x=0.66$ to $q_x=0.70$ it is possible to see the slope may be different than in the other part of the data set. When the lifetime goes beyond the ms, the explosive onset may occur too soon for the fit to properly render the first low temperature plateau.

\begin{table}
\centering
\begin{tabular}{|c|c|c|c|}
\hline 
N & $256$ & $512$ & $1024$ \\ 
\hline 
p & $-20.8$ & $-18.9$ & $-16.4$ \\ 
\hline 
r & $7.96\cdot 10^{-8}$ & $2.02\cdot 10^{-8}$ & $9.11\cdot 10^{-7}$ \\ 
\hline 
\end{tabular} 
\caption{Parameters for the fit of $t_0$ vs. $q_x$, for three ensembles of ions.}
\label{tab:t0vsqx_fit}
\end{table}

\subsection{Explosive onset temperature}
\begin{figure}[h]
  \includegraphics[width=0.44\textwidth]{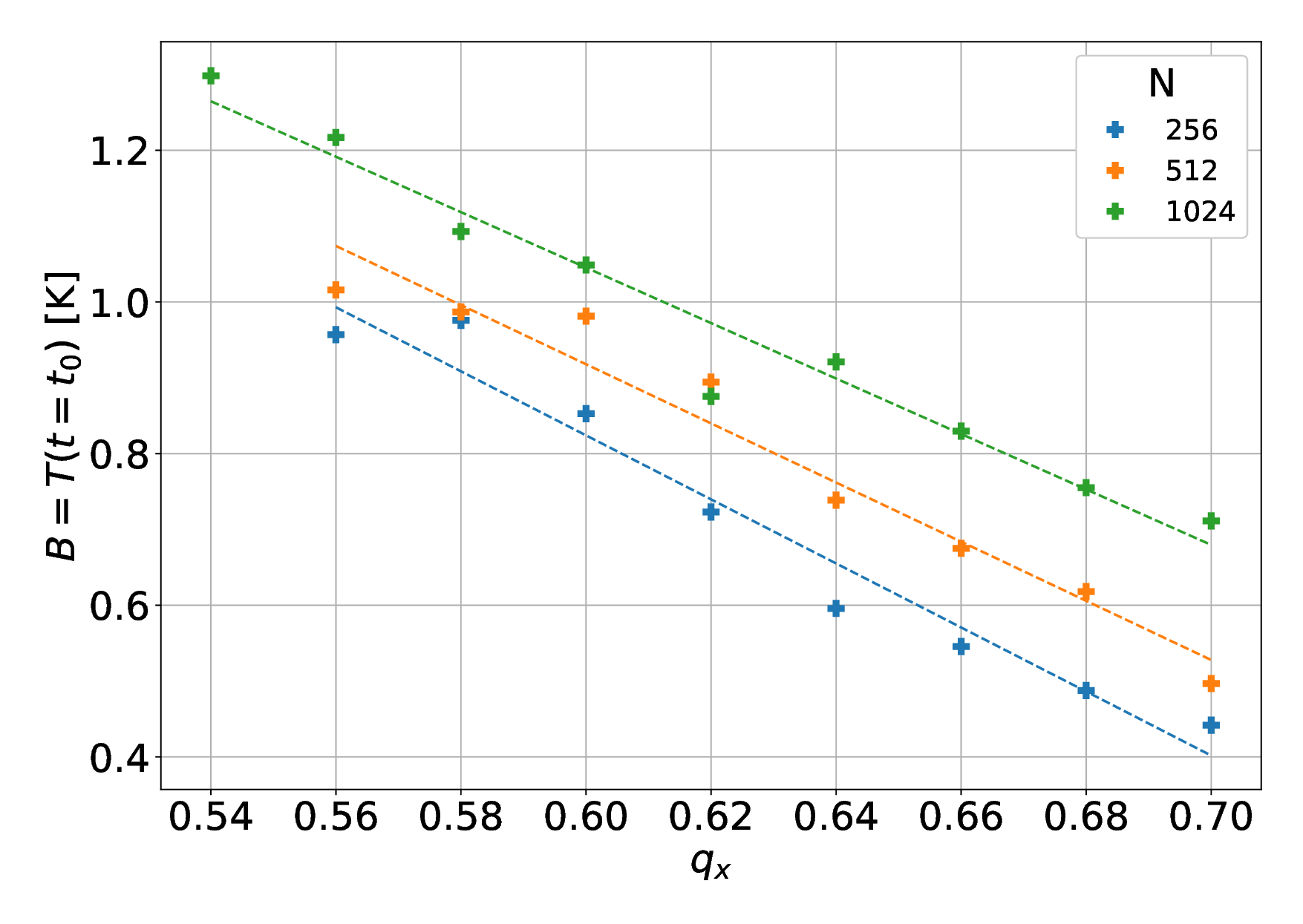}
  \caption{Cloud temperature $B$ at $t=t_0$ as a function of Mathieu parameter $q_x$. The dashed lines are affine fits to the corresponding data (colour code).}
  \label{fig:Bvsqx}
\end{figure}

The temperature at which the explosive onset occurs is given by the parameter $B=T(t=t_0)$, presented in figure \ref{fig:Bvsqx}. Globally the temperature of the explosive onset decreases linearly with $q_x$ in a linear. In the context of our trapped ensemble used as a molecular ion detector \cite{poindron21}, this temperature can be related to the trigger threshold of the detector. Indeed, once the ensemble overcomes this temperature, the rf heating becomes dominant with respect to laser cooling \cite{poindron23} and temperature reaches much higher orders of magnitude, providing a detection signal in the fluorescence of the ensemble \cite{poindron21}. This trigger temperature can be then converted into energy transferred from the projectile to the cloud. For instance, in a $N=1024$ \ce{^40Ca+} ion ensemble, an explosive onset temperature of $B=1$ K is equivalent to an energy of $132$ meV.

\subsection{Non-linear behaviour}

\begin{figure}[h]
  \includegraphics[width=0.44\textwidth]{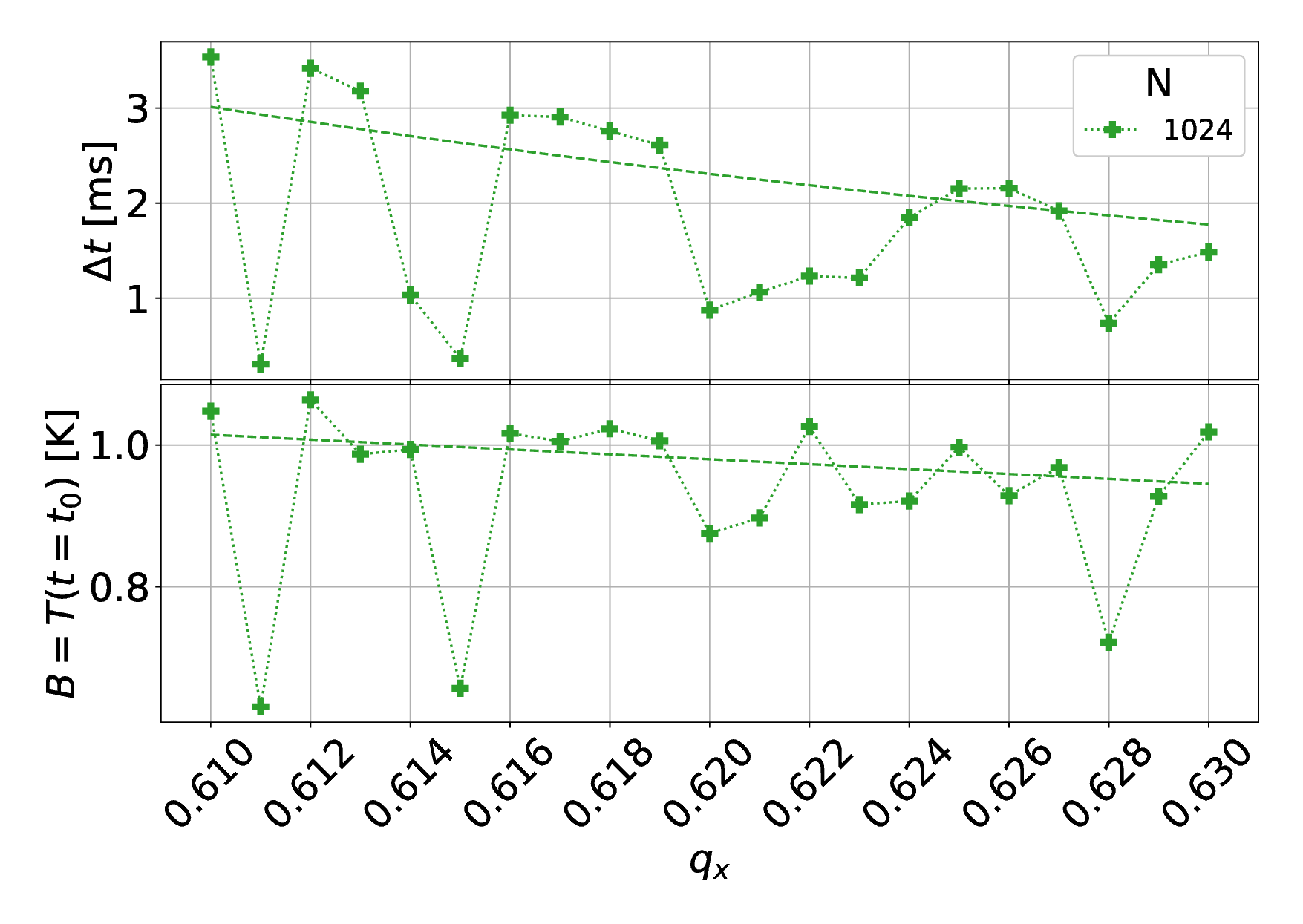}
  \caption{(Top) Ensemble lifetime $\Delta t$ as a function of Mathieu parameter $q_x$ for values around $q_x=0.62$. The dashed line is the fit from figure \ref{fig:t0vsqx}. The dotted line is a guide to the eye. (Bot) Explosive onset temperature $B$ as a function of $q_x$ for the same values around $q_x=0.62$. The dashed line is the fit from figure \ref{fig:Bvsqx}. The dotted line is a guide to the eye.}
  \label{fig:t0vsqx_zoom}
\end{figure}

At $q_x=0.62$ for $N=1024$ ions, it is possible to see that the ensemble lifetime before explosion is a third of the lifetime for similar conditions in the vicinity. At such parameters the temperature of the explosive onset is 10\% lower than for close $q_x$. In the context of our detector this is of particular interest because it allows the detector to be more sensitive. Additionnal simulation between $q_x=0.61$ and $q_x=0.63$ are carried out (Fig. \ref{fig:t0vsqx_zoom}). It shows that this extra sensitivity is achieved for $q_x=0.62\pm 0.01$, where $q_x$ only varies by 2\% in relative value. This margin is very wide regarding the precision of radio-frequency drives, thus it is possible to set the parameters in this area in order to make the ensemble more susceptible to a molecula ion. The existence of local conditions with significantly different outcomes in term of stability was already something visible in our previous study of the interaction of a trapped ensemble with a molecular ion \cite{poindron21}. It may be related to non-linear instabilities \cite{drakoudis06}. In this case, the explosive onset having a lower temperature as around allows to trigger more easily the detection with an injected molecular ion.

\section{Conclusion}

Typically, the fit provides a very consistent information on the explosive onset timing ($t_0$). By carefully avoiding to include parts of the temperature data where the evolution do not account for the radio-frequency heating we are studying, it is also possible to benefit from the information about the temperature at which explosive onset occurs ($B$). Ideally, an analytical formula derived from the microscopic understanding of the ensemble behaviour will also provide the same informations, but the determination of such time-dependent formula for the temperature evolution is a challenge \cite{tarnas13,nam17}, and answering to this problem requires a microscopical theory that our synthetical approach can't elude. Perhaps it is possible to elaborate a more representative sigmoid function once derived provides a more representative heating rate than the one we propose. Nevertheless, our method already provides a very good estimate of the lifetime of the cloud, which is a very crucial information for some experiments where trapped ions must be kept stable in cold temperature while cooling is sporadically interrupted.


%

\end{document}